\documentclass[aps,pra,twocolumn,superscriptaddress,amsmath,amssymb,showpacs]{revtex4}
\usepackage{graphicx}
\usepackage[T1]{fontenc}
\usepackage{textcomp}
\usepackage{color}

\begin{document}

\title{Suppression of non-adiabatic losses of molecules from chip-based microtraps}
\author{Samuel A. Meek}
\email{meek@fhi-berlin.mpg.de}
\author{Gabriele Santambrogio}
\affiliation{Fritz-Haber-Institut der Max-Planck-Gesellschaft,
Faradayweg 4-6, 14195 Berlin, Germany}
\author{Boris G. Sartakov}
\affiliation{A.M. Prokhorov General Physics Institute, RAS, Vavilov Street 38, Moscow 119991, Russia}
\author{Horst Conrad}
\author{Gerard Meijer}
\affiliation{Fritz-Haber-Institut der Max-Planck-Gesellschaft,
Faradayweg 4-6, 14195 Berlin, Germany}
\date{\today}

\begin{abstract}
Polar molecules in selected quantum states can be guided, decelerated, and trapped 
using electric fields created by microstructured electrodes on a chip. Here we explore 
how non-adiabatic transitions between levels in which the molecules are trapped and
levels in which the molecules are not trapped can be suppressed. We use $^{12}$CO
and $^{13}$CO ($a\, ^3\Pi_1, v=0$) molecules, prepared in the upper $\Lambda$-doublet
component of the $J=1$ rotational level, and study the trap loss as a function of an 
offset magnetic field. The experimentally observed suppression (enhancement) of the
non-adiabatic transitions for $^{12}$CO ($^{13}$CO) with increasing magnetic field is
quantitatively explained. 
\end{abstract}
\pacs{37.10.Pq, 31.50.Gh, 37.10.Mn}

\maketitle

\section{Introduction}
The manipulation and control of polar molecules above a chip using electric fields 
produced by microstructured electrodes on the chip surface is a fascinating new 
research field \cite{meek09a}. Miniaturization of the electric field structures enables 
the creation of large field gradients, i.e., large forces and tight potential wells for polar 
molecules. A fundamental assumption that is made when considering the force imposed 
on the molecules is that their potential energy only depends on the electric field strength. 
This is usually a good assumption, since the molecules will reorient themselves and 
follow the new quantization axis when the field changes direction and their potential energy 
will change smoothly when the strength of the field changes. This approximation can 
break down, however, when the quantum state that is used for manipulation couples to 
another quantum state that is very close in energy. If the energy of the quantum state 
changes at a rate that is fast compared to the energetic splitting, transitions between 
these states are likely to occur. For trapped molecules in so-called low-field-seeking 
states in a static electric potential, such transitions are particularly disastrous when they end 
up in high-field-seeking states or in states that are only weakly influenced by the electric fields, 
as this results in a loss of the molecules from the trap. This effect has been investigated 
previously for ammonia molecules in a Ioffe-Pritchard type electrostatic trap with a variable 
field minimum. In this macroscopic electrostatic trap, losses due to non-adiabatic transitions
were observed on a second time scale when the electric field at the center of the trap was 
zero; with a non-zero electric field minimum at the center of the trap, these losses could be 
avoided \cite{kirste09}. Non-adiabatic transitions have recently also been investigated in a 
``conventional'', i.e.\ macroscopic, Stark decelerator in which electric fields are rapidly 
switched between two different configurations. There, these transitions have been found to 
lead to significant losses of molecules when they are in low electric fields \cite{wall10}.
Similar trap losses will be much more pronounced on a microchip, where the length scales 
are much shorter and where the electric field vectors change much faster. For atoms in a 
three-dimensional magnetic quadrupole trap, the trap losses due to spin flip (or Majorana)
transitions has been shown to be inversely proportional to the square of the diameter of 
the atom cloud \cite{petrich95}. On atom-chips,
where paramagnetic atoms are manipulated above a surface using magnetic fields produced
by current carrying wires, trap losses due to Majorana transitions are therefore well-known but
can be conveniently prevented by using an offset magnetic field \cite{fortagh07}.
Due to the geometry of the molecule chip, however, applying a static offset electric field 
is not possible, and other solutions must be sought.

We have recently demonstrated that metastable CO molecules, laser-prepared in the upper
$\Lambda$-doublet component of the $J=1$ level of the $a\, ^3\Pi_1, v=0$ state can be guided,
decelerated and trapped on a chip. In these experiments, non-adiabatic losses have been
observed for $^{12}$C$^{16}$O. In this most abundant carbon monoxide isotopologue, the 
level that is low-field-seeking becomes degenerate with a level that is only weakly influenced 
by an electric field when the electric field strength goes to zero. Every time that the trapped 
molecules pass near the zero field region at the center of a micro-trap, they can make a transition 
between these levels and thereby be lost from the trap. This degeneracy is lifted in 
$^{13}$C$^{16}$O due to the hyperfine splitting (the $^{13}$C nucleus has a nuclear spin 
$|\vec I| = 1/2$), and the low-field-seeking levels never come closer than 50~MHz to the non-trappable levels. 
Therefore, changing from $^{12}$C$^{16}$O to $^{13}$C$^{16}$O (referred to as $^{12}$CO 
and $^{13}$CO from now on) in the experiment greatly 
improves the efficiency with which the molecules can be guided and decelerated over the 
surface and enables trapping of the latter molecules in stationary traps on the chip \cite{meek09a}. 

Although it is evident that the 50~MHz splitting between the low-field-seeking and non-trapped 
levels in $^{13}$CO is beneficial, it is not \emph{a priori} clear whether a smaller splitting 
would already be sufficient or if a still larger splitting would actually be needed to prevent all losses.
While the hyperfine splitting in $^{13}$CO cannot be varied, the degeneracy can be lifted 
by a variable amount in the normal $^{12}$CO isotopologue by using a magnetic field.
If a magnetic field is applied in addition to the electric field, a splitting can be induced between 
the low-field-seeking and non-trappable levels of $^{12}$CO that depends on the strength 
of the applied magnetic field; in $^{13}$CO, a magnetic field will actually decrease the splitting between 
the low-field-seeking and non-trappable levels.


In this paper, we present measurements of the efficiency with which CO molecules are transported
over the chip --- while they are confined in electric field minima that are traveling at a 
constant velocity --- as a function of magnetic 
field strength. It is observed that, in the case of $^{12}$CO, the losses due to non-adiabatic 
transitions can be completely suppressed in sufficiently high magnetic fields; for $^{13}$CO on
the other hand, the trap losses increase with increasing magnetic field. A theoretical model that 
can quantitatively explain these observations is also presented. Although the chip is ultimately 
used to decelerate molecules to a standstill, measuring the efficiency with which molecules are 
guided at a constant velocity provides a detailed insight into the underlying trap-loss mechanism. 
Limiting the experiments to constant velocity guiding also makes them more tractable:
bringing molecules to a standstill and subsequently detecting them has thus far required five separate 
phases of acceleration, which greatly complicates efforts to understand the details of the loss mechanism 
using numerical calculations \cite{meek09a}. While deceleration at a constant rate to a non-zero final 
velocity is possible, the measurable signal in such experiments is significantly lower than in constant 
velocity guiding. Despite the fact that we only measure at constant velocity, we nonetheless apply the 
model to examine the non-adiabatic losses expected during deceleration. This work thereby furthers the 
goal of extending trapping on the molecule chip to a wider range of molecules.

\section{Experimental setup}\label{expsetup}

\begin{figure}
\center{\includegraphics{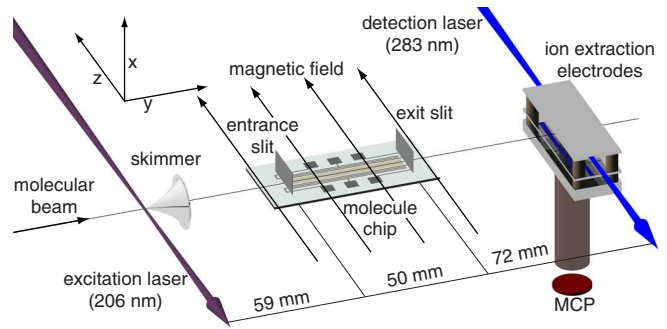}}
\caption{Schematic of the experimental setup. The laser prepared metastable CO molecules are
guided over the chip in tubular electric field minima that move at a constant speed of 300~m/s.
Above the chip, the molecules are exposed to a time-dependent electric field as well as to a constant
magnetic field whose direction is always perpendicular to the electric field. The guided CO
molecules are ejected from the chip, ionized via laser induced resonance enhanced multi photon ionization
(REMPI), and the CO$^+$ ions are mass-selectively detected on a micro-channel plate (MCP) 
detector. The coordinate system used in the further description is explicitly indicated.}
\label{beamline}
\end{figure}

A scheme of the experimental setup is shown in Figure~\ref{beamline}. A mixture of 20\% CO in krypton 
is expanded into vacuum from a pulsed valve (General Valve, Series 99), cooled to a temperature of 
140~K. In this way, a molecular beam with a mean velocity of 300~m/s and with a full-width-half-maximum 
spread of the velocity distribution of  approximately 50~m/s is produced. This beam passes through two 
1~mm diameter skimmers and two differential pumping stages (the valve and first skimmer are not shown 
in the figure) before entering the chamber in which the molecule chip is mounted. Just in front of the second 
skimmer, the ground state CO molecules are excited to the upper $\Lambda$-doublet component of the 
$J=1$ level of the metastable $a\, ^3\Pi_1$, $v = 0$ state, using narrow-band pulsed laser radiation 
around 206~nm (1~mJ in a 5~ns pulse with a bandwidth of about 150~MHz). The metastable CO 
molecules are subsequently guided in traveling potential wells that move at a constant speed of
300~m/s parallel to the surface of the molecule chip. A uniform magnetic field is applied to the region 
around the chip using a pair of 30~cm diameter planar coils separated by 23~cm (not shown in the figure).  
The coils are oriented such that the magnetic field is parallel to the long axis of the chip electrodes, i.e.\ 
along the $z$-axis, ensuring that the magnetic field is always perpendicular to the electric field (\emph{vide infra}).
The CO molecules that have been stably transported over the chip will pass through the $50$~\textmu m high 
exit slit and enter the ionization detection region a short distance further downstream. There, the
metastable CO molecules are resonantly excited to selected rotational levels in the $b\, ^3\Sigma^+$, 
$v'=0$ state using pulsed laser radiation at 283~nm (4~mJ in a 5~ns pulse with a 0.2~cm$^{-1}$ 
bandwidth). A second photon from the same laser ionizes the molecules and the parent ions are 
mass-selectively detected in a compact linear time-of-flight setup using a micro-channel plate (MCP)
detector. This detection scheme has been implemented in addition to the Auger detection scheme
that we have used in earlier studies \cite{meek09a,meek08,meek09} as it is more versatile and can 
also be applied to detect other molecules. In addition, the detection sensitivity of the ion detector 
is less affected by the magnetic field than that of the Auger detector. 

\begin{figure}
\center{\includegraphics{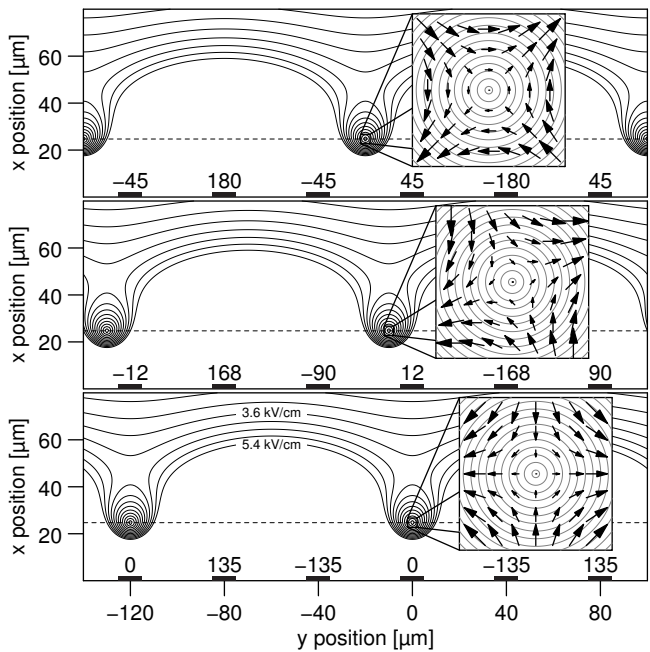}}
\caption{Edge-on view (in the $+z$-direction) of the molecule chip with calculated contour lines of 
equal electric field strength above the chip, displayed at intervals of 0.45~kV/cm. The three panels 
correspond to three different times in the harmonic waveform cycle. The position of the electrodes is 
indicated at the bottom of each panel and the instantaneous values of the applied potentials (in volts) are 
given. The time difference between adjacent panels is $\frac{1}{12\nu}$, where $\nu$ is the frequency 
of the harmonic waveforms. From the inset in each panel, it is seen that the electric field at each position
relative to the center of the trap rotates clockwise over an angle of $\frac{\pi}{4}$ between adjacent panels.}
\label{electrodes}
\end{figure}

The molecule chip and its operation principle have been described in detail before \cite{meek08,meek09}, 
and only the features that are essential for understanding of the present experiment are discussed 
here. The active area of the chip consists of an array of 1254 equidistant electrodes, each 10~\textmu m wide 
and 4~mm long, with a center-to-center distance of 40~\textmu m. An edge-on view of the chip electrodes (with 
the 4~mm dimension of the electrodes perpendicular to plane of the figure) is shown in Figure~\ref{electrodes}.
The potential (in volts) applied to an electrode at a given moment in time is indicated directly above the
electrode; these six potentials are repeated periodically on the electrodes on either side of those drawn 
here. Because the electrodes are much longer than the period length of the array, the electric field 
distribution can be regarded as two-dimensional, i.e.\ the component of the electric field along the $z$-axis 
can be neglected. This is of importance for the present experiments, because only in this case the applied
magnetic field is always perpendicular to the electric field. The calculated contour lines of equal electric 
field strength in the free space above the chip 
show electric field minima that are separated by 120~\textmu m, i.e., there are two electric field minima 
per period, centered about 25~\textmu m above the surface of the chip. By applying sinusoidal 
waveforms with a frequency $\nu$ to the electrodes, these minima can be translated parallel 
to the surface with a speed given by $v = 120\, \textrm{\textmu m}\cdot \nu$. When these 
waveforms are perfectly harmonic and have the correct amplitude, offset, and phase, the minima move 
with a constant velocity at a constant height above the surface, and the shape of the field strength 
distribution does not change in time. Because an electric field strength minimum acts as a trap for 
molecules in low-field-seeking states, these fields act as tubular moving traps that can be used to 
guide the molecules over the surface of the chip. The tubular traps are closed at the end by the fringe 
fields caused by the neighboring electrodes. Near the ends of the about 4~mm long traps, the electric 
field will necessarily have a component along the $z$-axis, i.e.\ parallel to the applied magnetic field. 
In the present study, where the molecules are guided at 300~m/s over the chip and are therefore on 
the chip for less than 200~\textmu s, these end-effects are neglected.

The region near an electric field minimum at $(x,y) = (x_0,y_0)$ is a quadrupole, with an electric 
potential given by
\begin{equation}
V = \frac{\alpha}{2} r^2 \cos(2 \phi - \phi_0)
\end{equation}
where $x - x_0 = r \cos\phi$ and $y - y_0 = r \sin\phi$. In the current experiment, sinusoidal waveforms 
with an amplitude of 180~V are applied to the electrodes, yielding a value of 
$\alpha = 0.054\, \textrm{V}/\textrm{\textmu m}^2$ \cite{meek09}.
The resulting electric field is given by
\begin{equation}
\vec{E} = -\alpha r (\cos(\phi_0 - \phi) \hat{x} + \sin(\phi_0 - \phi) \hat{y}).
\end{equation}
The strength of the electric field, $|\vec{E}| = \alpha r$, depends only on the $r$ coordinate, but 
the direction of the field vector, $\phi' = \phi_0 - \phi + \pi$, depends on the coordinate $\phi$ and on 
the phase factor $\phi_0$. While the direction of the field vector changes as a result of the motion of the 
molecule in the quadrupole field (and thus changing $\phi$), the direction of the field at any given 
position relative to the minimum also rotates when the minimum is translated over the chip. It is 
seen from the electric field vectors shown in the insets of Figure~\ref{electrodes} that the frequency 
of this rotation is 1.5 times the frequency of the applied waveforms and that the direction of the rotation
is clockwise, i.e.\ the rotation vector points along the positive $z$-axis and $\phi_0$ increases 
linearly in time. To guide the molecules over the chip at 300~m/s, harmonic waveforms with a 
frequency $\nu$ of 2.5~MHz must be applied, resulting in a rotation frequency of 3.75~MHz. 

For CO molecules in the low-field-seeking component of the $J=1$ level, the depth of the tubular
traps above the chip is about 60~mK. This implies that CO molecules with a speed of up to 
6~m/s relative to the center of the trap can be captured. The oscillation frequency of the
molecules in the radial direction of the tubular traps is in the 100--250~kHz range. These parameters 
are of importance for the non-adiabatic transitions as these determine with which velocity, 
and how often per second the CO molecules pass by the zero-field region of the traps. 

\begin{figure}
\center{\includegraphics{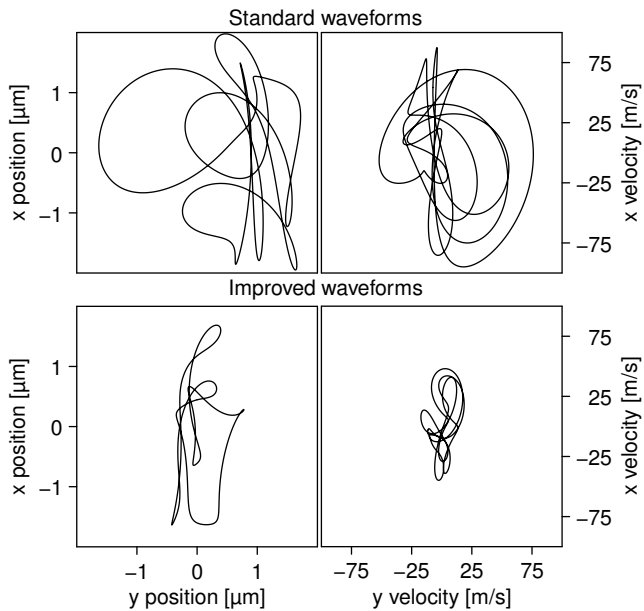}}
\caption{Calculated position of the minimum of an electric field trap with 
respect to an ideal trap moving at constant velocity during an 800~ns 
time-interval together with the resulting velocity in the $xy$ plane, using the measured 
``standard waveforms'' (upper row) and ``improved waveforms'' (lower row) as input.}
\label{jitter}
\end{figure}

It turns out that, in the actual experiment, the tubular traps do not move perfectly smoothly
over the chip. Due to imperfections in the amplitude, offset and phase of the waveforms 
that we have used, the tabular traps are jittering at rather high velocities. The motion of 
the center of the traps relative to the ideal, constant velocity motion can be 
calculated by measuring the real waveforms applied to the chip and using these to compute 
the position of the minimum for each point in time. The upper row of Figure \ref{jitter} shows 
the motion of a minimum when the waveforms that we will refer to as the ``standard waveforms'' 
have been used. The range of this motion extends over $\pm2$~\textmu m in the $x$ and $y$ 
coordinates. Although this is considerably smaller than the size of the trapping region, 
this motion significantly enlarges the effective region in which non-adiabatic transitions 
can occur. Moreover, the entire path is traced out periodically every $T = 2/\nu = 800$~ns;
because the motion occurs on such a short time scale, 
the speed with which the trap center moves, and therefore the relative speed with which the 
molecules encounter the trap center, can be as high as 100~m/s. To improve the waveforms, we inserted 
an LC filter in the output stage of the amplifiers, thereby reducing the harmonic distortion. 
The resulting motion using these ``improved waveforms'' is shown in the lower row of 
Figure~\ref{jitter}. It is seen that not only the range of the motion is now contracted but that
also the speed with which the trap center jitters is reduced by about a factor of two.  

\section{Theoretical Model}\label{theo}
\subsection{Eigenenergies in combined fields}
In order to describe the non-adiabatic transitions in CO, we must first derive the
energy levels of CO in combined electric and magnetic fields. For this, the field-free 
Hamiltonian is expanded with Stark and Zeeman contributions, i.e., 
\begin{equation}
\hat{H}=\hat{H}_{\Lambda,\textrm{hfs}}+\hat{H}_{S}+\hat{H}_{Z}.
\end{equation}
Here, $\hat{H}_{\Lambda,\textrm{hfs}}$ describes the $\Lambda$-doubling of
the $a\, ^3\Pi_1$, $v=0$, $J=1$ level for either $^{12}$CO or $^{13}$CO and also includes
the hyperfine splitting of each $\Lambda$ component into $F=1/2$ and $F=3/2$
hyperfine sublevels for $^{13}$CO;
$\hat{H}_{S}=-\hat{\vec{\mu}}_E\cdot\vec{E}$ is the Stark interaction
Hamiltonian and $\hat{H}_{Z}=-\hat{\vec{\mu}}\cdot\vec{B}$ is the Zeeman
interaction Hamiltonian, where $\hat{\vec{\mu}}_E$ and
$\hat{\vec{\mu}}$ are the electric and magnetic dipole moment operators,
respectively, and $\vec{E}$ and $\vec{B}$ are the (time-dependent) electric 
and magnetic field vectors.

The spectroscopic parameters of the $a\, ^3\Pi_1$, $v=0$ state of CO
that are used in the field-free Hamiltonian are given 
elsewhere \cite{meek10,wicke72,gammon71,carballo88,yamamoto88,wada00,field72,warnerphd}.
The $\Lambda$-doublet splitting between the positive parity component (upper) 
and the negative parity component (lower) of the $J=1$ level is about 400~MHz 
while the hyperfine splitting of each parity level of $^{13}$CO into $F=1/2$ (lower) 
and $F=3/2$ (upper) sublevels is about one order of magnitude smaller. The
body-fixed electric dipole moment $\mu_E = |\vec{\mu}_E|$ in the electronically 
excited metastable state is 1.3745 Debye for both $^{12}$CO and
$^{13}$CO \cite{wicke72,gammon71}. The magnetic moment of the 
molecule can be expressed as $\hat{\vec{\mu}}=-\mu_B\,(g_L\cdot \hat{\vec{L}} +
g_S\cdot \hat{\vec{S}})$, where $\mu_B$ is the Bohr magneton, $\hat{\vec{L}}$ is the 
electron orbital angular momentum operator, $\hat{\vec{S}}$ is the electron spin 
operator, and where the magnetic $g$-factors are fixed at the values of the bare 
electron, $g_L =1.0$ and $g_S = 2.0023$.

\begin{figure}
\center{\includegraphics{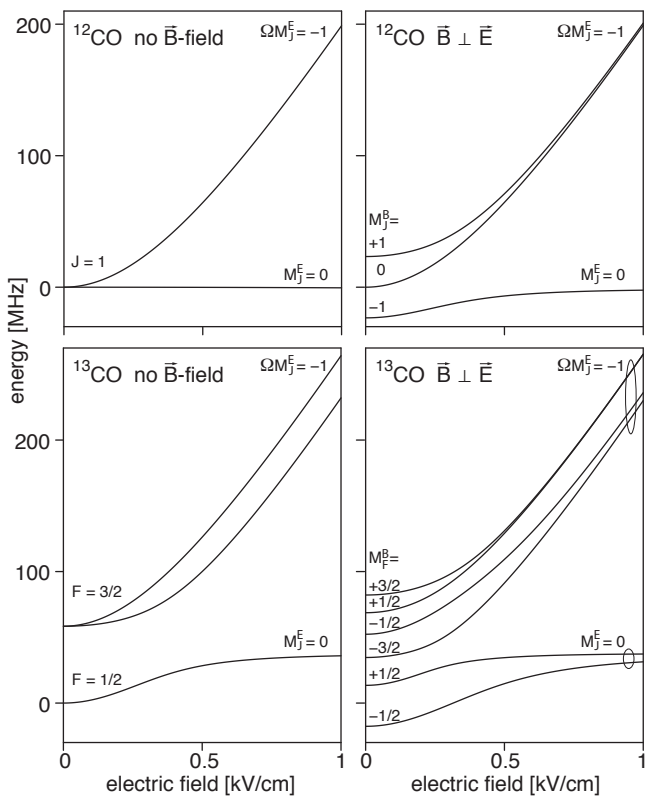}}
\caption{Energy of the upper (positive parity) $\Lambda$-doublet component of the 
$J=1$ level of $^{12}$CO (upper panels) and $^{13}$CO (lower panels) in the electronically 
excited $a\, ^3\Pi_1$, $v=0$ state as a function of the electric field strength, without 
(left column) and with (right column) an offset magnetic field. The strength of the 
magnetic field is 50~Gauss and its direction is perpendicular to the direction of the 
electric field. The labeling of the energy levels is explained in the text.}
\label{colevel}
\end{figure}

A detailed description of the formalism used to calculate the eigenenergies of the 
various components of the $J=1$ level in the $a\, ^3\Pi_1$, $v=0$ state of both 
$^{12}$CO  and $^{13}$CO in combined electric and magnetic fields is presented 
in the Appendix. The formalism has been set up for mutually orthogonal static electric 
and magnetic fields. As will be discussed below, this is also adequate to treat the actual 
situation, in which the electric field rotates with a constant frequency in a plane 
perpendicular to the magnetic field. In Figure~\ref{colevel} we only show the 
outcome of these calculations in the form of plots of the energy levels for the upper
$\Lambda$-doublet component of $^{12}$CO and $^{13}$CO as a 
function of electric field strength in the absence (left column) or in the presence of 
a 50~Gauss magnetic field (right column). 
At low electric field strengths, the Stark shift is quadratic, but as the electric field 
strength increases and $|H_S|$ becomes much larger than $|H_{\Lambda,\textrm{hfs}}|$ and 
$|H_Z|$, the Stark energy shows a linear dependence on the electric
field strength, i.e.\ $\Delta E_{S} \propto -\Omega M_J^E \mu_E E$, and the 
product of the projection of the electronic angular momentum along the internuclear axis 
$\Omega$ and the projection $M_J^E$ of the angular momentum $\hat{\vec{J}}$ on the 
electric field vector $\vec{E}$ becomes an approximately good quantum number. 
If a weak magnetic field such as that present in the experiment is applied in the
absence of an electric field, each of the zero-field levels splits into 
$(2J + 1)$ (or in $(2F + 1)$) separate levels based on their $M_J^B$ (or $M_F^B$) 
quantum number. The Zeeman energy is linear in magnetic field and can be 
calculated using first order perturbation theory as 
$\Delta E_Z \approx \mu_{\textrm{eff}} M B$ where 
$M=M_J^B$ (or $M_F^B$) and $\mu_{\textrm{eff}}\sim \mu_B$ describes the effective 
magnetic moment of a particular parity and, in the case of $^{13}$CO, particular 
$F$ component. When a strong electric field is applied to the molecule in addition 
to the magnetic field, such that $|H_S| \gg |H_{\Lambda,\textrm{hfs}}|,|H_Z|$, 
the energy levels can again be characterized with the approximately good quantum 
number $\Omega M_J^E$ and show a linear Stark shift.

The behavior of the eigenenergies in static electric and magnetic fields is quite 
instructive in describing the non-adiabatic transitions --- and thus losses --- of 
molecules from the low-field-seeking states ($\Omega M_J^E = -1$) to non-trappable ($M_J^E = 0$) 
states. Near the edge of the trap, where the electric field is as high as 4.2~kV/cm,
the Stark effect provides an energy gap of about 
$U_{\textrm{trap}}= 60\, \textrm{mK}\simeq 1.3\, \textrm{GHz}$ between low-field-seeking states and 
non-trappable states. Since this is much larger than both the frequency of the motion 
of the molecules in the traps and the frequency of the applied waveforms, non-adiabatic
losses will not occur near the edge of the trap. In the vicinity of the trap center, however, 
this argument no longer holds, and the eigenfunctions can change from $M_J^E$-type 
wavefunctions to $M_J^B$-type (or $M_F^B$-type) at a rate faster than the energy 
gap $E_{\textrm{gap}}$ in that region. In the case of a $^{12}$CO molecule, the 
energy gap at the center of the trap goes to zero in the absence of a magnetic field. 
If a molecule in a low-field-seeking state flies near the trap center with a velocity $v$ 
high enough, or with a distance of closest approach $b$ small enough, that the corresponding 
interaction time with the trap center $\tau=b/v$ no longer fulfills the adiabaticity 
condition, i.e.\ when the condition $E_{\textrm{gap}} \gg h/\tau$ no longer holds, then the 
probability of transitions to non-trappable states can become significant. In the 
absence of a magnetic field, $^{13}$CO molecules are much safer from such 
non-adiabatic losses due to the energy gap of 50~MHz between the $F=3/2$ level 
(which becomes low-field-seeking in an electric field) and the $F=1/2$ level (which 
correlates with non-trappable $M_J^E=0$ states).

\subsection{Calculating rates for non-adiabatic transitions}
The basic idea underlying the calculation of the non-adiabatic losses for the 
molecules in low-field-seeking states is that, since the transitions to non-trappable 
states happen primarily as the molecules pass the zero field region at the center 
of a microtrap, the overall loss probability can be estimated by first calculating the 
loss probability in a single pass. For simplicity, the trajectory of the CO molecules
is assumed to have a constant velocity; this is reasonable, since the forces on the 
molecules approach zero at the center of the trap due to the $\Lambda$-doubling. 
The transition probability $P_{i,j}(v, B, b)$ of a molecule making a transition from
a low-field-seeking state $i$ to a non-trappable state $j$ for a single pass by the trap 
center depends on the speed of the molecule relative to the center of the trap $v$, 
on the strength of the magnetic field $B$, and on the distance of closest approach $b$. 
Due to the rotation of the electric field, there is a difference between positive 
and negative values of $b$, which is why we do not refer to it as an ``impact parameter'' here.

To calculate the probability $P_{i,j}(v, B, b)$, the time-dependent Schr\"odinger 
equation, which describes the evolution of the quantum states, must be solved:
\begin{equation}
i \hbar \frac{\partial \psi}{\partial t} = \hat{H} \psi.
 \label{tds}
\end{equation}
This equation is solved numerically as an initial value problem on a set of coupled first-order 
ordinary differential equations using the basis vectors given in the Appendix. The Hamiltonian 
$\hat{H}$ depends on the electric field vector $\vec{E}$, which, for a molecule moving in the trap, 
is a function of both position $\vec{r}(t)$ (due to the inhomogeneous field distribution) and time 
$t$ (due to the rotation of the field vectors relative to the trap center), as discussed in 
Section \ref{expsetup}. The time dependent Hamiltonian that the molecule experiences is 
calculated by assuming a position in time given by $r(t) = (v t - x_{\textrm{init}}) \hat{x} + b \hat{y}$, 
which corresponds to the molecule moving with a constant velocity $v$ and approaching the trap 
center with a minimum distance $b$.  The variable $x_{\textrm{init}}$ must be chosen such that the initial 
position $r(0) = -x_{\textrm{init}} \hat{x} + b \hat{y}$ is sufficiently far from the center such that at 
$r(0)$ the adiabaticity condition is still satisfied. 

The initial state of the molecule could be chosen as a low-field-seeking eigenstate of the 
instantaneous Hamiltonian at the initial position and  time. If the electric field vector is rotating, 
however, the wavefunction of the molecule will immediately accumulate amplitude in other 
quantum states, even if the magnitude of the electric field is constant. The amplitude that 
appears in other states will only be negligible if the energy splitting between them and the initial 
state is much larger than the rotation frequency. Alternatively, one can choose an initial state that 
is a stationary state in the rotating system, as will be discussed below. The calculation of the 
transition probability as the molecule flies past the field minimum can then be started at lower fields, 
i.e.\ with a smaller value of $x_{\textrm{init}}$.

In the current system, the magnetic field vector is oriented along the $\hat{z}$ axis and the 
electric field vector lies in the $xy$ plane. If the angle of the electric field vector 
with respect to the $+\hat{x}$ axis (toward the $+\hat{y}$ axis) is given by $\phi'$, the 
Hamiltonian of the molecule can be computed by rotating the physical system around the 
$\hat{z}$ axis through an angle $-\phi'$, operating on it with a Hamiltonian $\hat{H}'$ which 
corresponds to a system with the same magnetic field vector and in which the electric field 
has the same magnitude but is directed along the $+\hat{x}$ axis, and then rotating the 
physical system back through an angle $\phi'$. In operator form, this can be written as
\begin{equation}
\hat{H} = e^{-i \hat{F}_z \phi'} \hat{H}' e^{i \hat{F}_z \phi'}
\label{HprimeH}
\end{equation}
where $\hat{F}_z$ is the total angular momentum of the molecule along the $\hat{z}$ axis.  
Using this form of $\hat{H}$, equation (\ref{tds}) can be rewritten as
\begin{equation}
i \hbar \frac{\partial \psi_Q}{\partial t} = \Bigl(\hat{H}' - \hat{F}_z \hbar \frac{\partial \phi'}{\partial t}\Bigr) \psi_Q
\label{qeh}
\end{equation}
where $\psi_Q = e^{i \hat{F}_z \phi'} \psi$.
This equation has the same form as equation (\ref{tds}) and can be solved in the same way 
as the time independent Schr\"odinger equation if the magnetic field is constant, the electric 
field strength is constant, and the electric field vector rotates at a constant frequency. The 
eigenvalues that result are called ``quasienergies'' and 
$\hat{H}_Q = \hat{H}' - \hat{F}_z \hbar \frac{\partial \phi'}{\partial t}$ 
is the quasienergy Hamiltonian \cite{zeldovich67,ritus67}. A similar approach has been used in 
the recent work by Wall et al.\ \cite{wall10}.

In section II it was shown that the angle of the electric field in the $xy$ plane 
$\phi' = \phi_0 - \phi + \pi$ contains contributions from both the angular coordinate of the 
molecule with respect to the trap center $\phi$ and a phase that results from the constant rotation
of the field vectors as the traps move over the chip $\phi_0 = 2\pi \frac{3\nu}{2}\,t$.
Because $\phi_0$ increases linearly in time, the contribution of $\phi_0$ to the quasienergy
Hamiltonian is time independent, having the form $-h \frac{3\nu}{2} \hat{F}_z$. This operator is
diagonal in the basis sets used for both $^{12}$CO and $^{13}$CO, and in the case of $^{12}$CO, it
is exactly equivalent to the Zeeman interaction for each energy eigenstate. As the rotation frequency 
$\frac{3\nu}{2}$ is 3.75~MHz when the molecules are guided with 300~m/s over the chip, 
its contribution to the quasienergy Hamiltonian is equivalent to a magnetic field of about $-8$~Gauss. 
It should be understood, however,  that we are not dealing with a real magnetic field produced by
the rotating electric field here; 
the effect of a rotating coordinate system merely produces a shift to the quasienergies that 
resembles the Zeeman Hamiltonian but that does not depend on the magnetic moment. 
In the case of $^{13}$CO, the rotation frequency is not equivalent to a Zeeman interaction
since, as stated in the definition of $\hat{\vec{\mu}}$, the gyromagnetic factors of the orbital 
angular momentum and the electron spin are different.

For the subsequent calculations, the quasienergy eigenvector instead of the normal Hamiltonian 
eigenvector is chosen as the initial state.  After choosing an initial state $i$ at the position 
$r(0) = -x_{\textrm{init}} \hat{x} + b \hat{y}$, the quasienergy vector is propagated in time using 
equation (\ref{qeh}) until the molecule reaches the position 
$r(t_\textrm{final}) = x_{\textrm{init}} \hat{x} + b \hat {y}$. 
The final state is then expressed in terms of quasienergy eigenvectors at the final position and time,
and the probability of the molecule ending up in a state $j$, $P_{i,j} (v, B, b)$, is calculated.
For all calculations, the population is assumed to be initially distributed equally over all 
low-field-seeking levels, two for $^{12}$CO and four for $^{13}$CO. In $^{12}$CO, the calculation of 
$P_{i,j} (v, B, b)$ only needs to be carried out for the initial state $i$ corresponding to the $M_J^B=+1$ 
low-field-seeking level, since the $M_J^B=0$ 
low-field-seeking level is completely stable against non-adiabatic transitions (see the Appendix).
For $^{13}$CO, none of the four low-field-seeking levels is stable against non-adiabatic transitions 
at all magnetic fields, so $P_{i,j} (v, B, b)$ must be calculated for each initial level $i$.

In the end, we are interested in the probability $T(B)$ of a molecule remaining in a 
low-field-seeking state for the duration of its time in the microtrap, as this is the quantity 
that is measured in the experiment. For $^{12}$CO molecules in the ideal case, i.e.\ when the 
electric field is perfectly perpendicular to the magnetic field and the traps move perfectly 
smoothly over the chip, the survival probability for a single molecule in a single state is given 
by the product of its survival probabilities after each individual encounter with the trap center.
The overall transmission probability $T(B)$ is then calculated by averaging this over both 
low-field-seeking states and over $N$ molecules.
\begin{equation}
\begin{split}
T(B) = \frac{1}{N} \sum_{n=1}^N \Biggl[
&\frac{1}{2} \prod_{k=1}^K P_{M_J^B=0,M_J^B=0}(v_{n,k}, B, b_{n,k}) +\\
&\frac{1}{2} \prod_{k=1}^K P_{M_J^B=+1,M_J^B=+1}(v_{n,k}, B, b_{n,k})
\Biggr]
\end{split}
\end{equation}
The total number of passes $K$ of each molecule and the speed $v_{n,k}$ and closest approach distance 
$b_{n,k}$ of the $k$th pass of the $n$th molecule are determined using simulations of the classical 
trajectory of a molecule in the trap, as described elsewhere \cite{meek09}.  Since the $M_J^B=0$ state
is completely stable, $P_{M_J^B=0,M_J^B=0}(v_{n,k}, B, b_{n,k}) = 1$, and the transmission probability $T(B)$ in 
the ideal case can never be less than $1/2$.  For $^{13}$CO, calculating $T(B)$ is somewhat
more complicated, since the $F=3/2, M_F^B = +3/2$ low-field-seeking level and the $F=3/2, M_F^B = -1/2$ 
low-field-seeking level are coupled, as are the $F=3/2, M_F^B = +1/2$ and $F=3/2, M_F^B = -3/2$ levels. 
During each encounter with the trap center, a molecule in a particular low-field-seeking state can transition 
not only to a non-trappable state but also to one other low-field-seeking state. To calculate the transmission
probability of a single $^{13}$CO molecule, the population in each low-field-seeking state after an 
encounter with the trap center is computed based on the population distribution before the encounter, 
and the total population still in a low-field-seeking state after the last pass is recorded.  As in $^{12}$CO, 
$T(B)$ is obtained by averaging the result of this calculation over a large number of molecules.

To accurately describe the experimental data, the theoretical calculations must be extended to include
the jittering motion of the traps and the nonperpendicularity of the electric and magnetic fields.
The jittering motion is accounted for by including the full motion of the center of the trap as 
shown in Figure~\ref{jitter} in the calculation of the transition probability $P_{i,j} (v, B, b, t, \phi)$. 
As in the ideal case, 
the transition probability depends on speed $v$, magnetic field $B$, and closest approach distance $b$ 
(although this distance is now defined relative to the average position of the minimum instead of the 
actual position). Additionally, the transition probability now also depends on the time $t$ at which 
the molecule arrives relative to the jittering cycle and the direction $\phi$ from which it comes. 
If the electric and magnetic fields are not exactly perpendicular, low-field-seeking states that are 
normally decoupled can mix.  In $^{12}$CO, the $M_J^B = 0$ and $M_J^B = +1$ states, which converge 
asymptotically at high electric fields, then become coupled.  As a result, 
the population can partially redistribute between these two levels while the molecule is in a region 
of high electric field between successive encounters with the trap center. To account for this effect, 
it is assumed in the calculations that, after each encounter with the trap center, a molecule's 
population $n$ in each of these two low-field-seeking states is redistributed such that 
\begin{align}
n'_{M_J^B = +1} &= (1-m) n_{M_J^B = +1} + m n_{M_J^B = 0}\\
n'_{M_J^B = 0} &= (1-m) n_{M_J^B = 0} + m n_{M_J^B = +1}
\end{align}
where $n'$ is the new population distribution. The parameter $m$ describes the degree of the 
redistribution; at the extremes, a value of $0$ indicates that no redistribution occurs while a value of 
$1/2$ corresponds to complete redistribution. Its exact value is difficult to predict and should actually 
depend on the trajectory of the molecule.  For simplicity, $m$ is determined by fitting it to the data; note 
that this is the only fitting parameter used. For $^{13}$CO, remixing can occur at high electric fields between 
the $F=3/2, M_F^B = -3/2$ and the $F=3/2, M_F^B = -1/2$ levels and also between the $F=3/2, M_F^B = +1/2$ and the 
$F=3/2, M_F^B = +3/2$ levels.  The remixing coefficient $m$ can be different for each of these pairs of levels, 
and thus for $^{13}$CO, two fitting parameters are necessary.

\section{Experimental results}
\begin{figure}
\center{\includegraphics{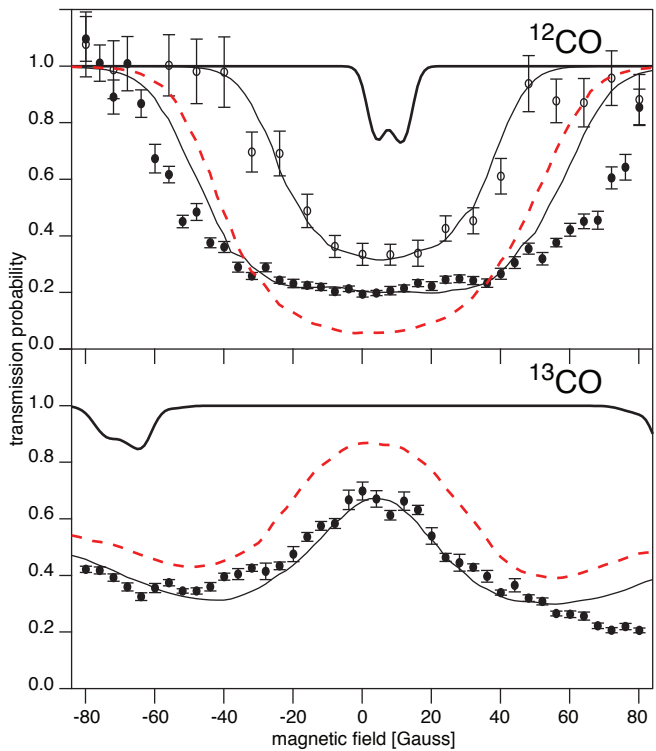}}
\caption{Measured signal of $^{12}$CO (upper panel) and $^{13}$CO (lower panel) 
molecules guided over the chip at 300~m/s as a function of the magnetic field. 
The vertical scaling of the individual data sets is described in the text.
Measurements have been performed using the ``standard waveforms'' (filled circles) 
for both $^{12}$CO and $^{13}$CO; for $^{12}$CO, measurements with the ``improved 
waveforms'' (open circles) are shown as well. The simulated curves of the transmission 
probability $T(B)$ are shown as solid curves that overlay the data points. The thick 
solid curves in the upper and in the lower panel result from the theoretical model for 
the ideal case. The red dashed curves show the prediction of the theoretical 
model for the survival probability of molecules decelerated from 300~m/s to zero 
velocity in 250~\textmu s.}
\label{data}
\end{figure}

To measure the effectiveness of the magnetic field at suppressing non-adiabatic 
losses, metastable CO molecules are guided over the full length of the chip at a 
constant velocity of 300~m/s and are subsequently detected using laser ionization.
Measurements are carried out for both positive and negative magnetic fields, where 
the direction of positive magnetic field coincides with the $+\hat{z}$-axis (see 
Figure~\ref{beamline}). To compensate for long term drifts in the intensity of the 
molecular beam, the parent ion signal is measured with the magnetic field on and 
with the magnetic field off, and the ratio between these two measurements is 
recorded. In Figure~\ref{data}, the thus recorded relative number of $^{12}$CO 
(upper panel) and $^{13}$CO (lower panel) molecules that are guided over 
the chip is shown as a function of the applied magnetic field. Both the 
measurements with the ``standard waveforms'' and with the ``improved 
waveforms'' are shown in the upper panel. 

It is clear from the data shown in Figure~\ref{data} that, as the magnetic field strength 
increases, the number of $^{12}$CO molecules reaching the detector increases. This
is as expected because the splitting between the low-field-seeking levels of $^{12}$CO 
and the non-trappable level increases with the magnetic field strength, 
thereby increasingly suppressing non-adiabatic losses. The measurements with the 
``improved waveforms'' show that at magnetic fields more negative than $-40$~Gauss 
and more positive than $+50$~Gauss, the number of guided molecules becomes constant, 
indicating that all non-adiabatic losses are suppressed under these conditions. The
data have been scaled vertically such that the saturation observed at high magnetic field
strengths corresponds to a transmission of unity.
The ratio between the signal at high magnetic field and low magnetic field 
is smaller than when the ``standard waveforms'' are used, implying that the ``improved 
waveforms'' also reduce the losses without a magnetic field. The number of $^{13}$CO 
molecules reaching the ionization detector, on the other hand, is seen to decrease as a 
function of magnetic field. The vertical scale in this case is based on the results of 
the theoretical calculations at low magnetic field strengths (\emph{vide infra}).
The lower panels of Figure~\ref{colevel} show that the 
$M_F^B=+1/2$ level of the non-trappable $F = 1/2$ state increases in energy in a magnetic 
field, while the $M_F^B=-3/2$ and $M_F^B=-1/2$ levels of the low-field-seeking $F = 3/2$ 
state are at the same time lowered in energy. This reduction of the energy gap between the 
low-field-seeking and non-trappable levels enhances the non-adiabatic losses. Both the 
$^{12}$CO and $^{13}$CO data are slightly asymmetric for positive and negative magnetic 
fields. The data for $^{12}$CO indeed seem to be symmetric around a magnetic field 
of about $+8$~Gauss, as expected from the theoretical model.

The results of the theoretical models are shown as solid curves in Figure~\ref{data} as 
well. The thick solid curves in the upper and in the lower panel result from the theoretical 
model for the ideal case, i.e.\ when the jittering of the traps would be absent and no remixing
would occur between decoupled states. The width of the theoretically predicted transmission minimum 
for $^{12}$CO around $+8$~Gauss depends sensitively on the relative velocity between the CO molecule 
and the center of the trap as they pass, and gets larger with increasing velocities. In the case 
of $^{13}$CO, two narrow transmission minima are expected around $-65$~Gauss and $+90$~Gauss, 
corresponding to fields at which the the $F = 1/2, M_F^B = +1/2$ and $F = 3/2, M_F^B = -3/2$ 
levels cross, and between these two minima, no losses are expected. As in $^{12}$CO, the 
width of the transmission minima increases as the relative velocity between the molecules and 
the trap center increases. It is clear from the comparison of these theoretical curves with 
the experimental data, that the observed measurements can not be quantitatively explained if the 
traps are assumed to move smoothly over the chip; the jittering of the traps must be taken 
into account.

The transmission probabilities calculated for the case in which the jittering is explicitly taken 
into account are seen to almost quantitatively agree with the measurements. In particular the
theoretical curves reproduce the asymmetry between the intensity of the guided $^{12}$CO 
molecules at positive and negative magnetic fields as well as the narrowing of the profiles when 
the waveforms are improved. In the calculations for $^{12}$CO, a partial redistribution after each pass 
of 18\% ($m = 0.09$) has been assumed for the ``standard waveforms'' and 12\% ($m = 0.06$) for the 
``improved waveforms''.  It can not be excluded that the jittering of the traps was still slightly 
more severe during the actual experiments than shown for the ``standard waveforms'' in 
Figure~\ref{jitter}, which would explain the experimentally observed additional broadening for 
that case. In the $^{13}$CO calculations, the best agreement with the experimental data was 
found when assuming no remixing between the $M_F^B = +3/2$ and the $M_F^B = +1/2$ levels, and 
30\% remixing ($m = 0.15$) between the $M_F^B = -1/2$ and the $M_F^B = -3/2$ level. The maximum 
transmission at about $+10$~Gauss is not sensitive to the remixing coefficients, since the 
transition probability for each of the low-field-seeking states is about the same in this region. 
It can be reliably inferred from the theoretical calculations that, even at low magnetic fields, 
about $1/3$ of the $^{13}$CO molecules are lost to non-adiabatic transitions while being guided 
over the chip. Based on this, the $^{13}$CO data shown in the lower panel of Figure~\ref{data}
have been scaled vertically such that the transmission at zero magnetic field is 2/3.

The theoretical model used to explain the guiding data can also be applied to predict 
the non-adiabatic losses that are expected to occur during linear deceleration. The red dashed 
curves in the upper (lower) panel of Figure~\ref{data} show the survival probability of 
$^{12}$CO ($^{13}$CO) molecules decelerated from 300~m/s to zero velocity in 250~\textmu s. In 
these calculations, it is assumed that the jittering motion at 300~m/s is that of the ``standard 
waveforms''. The velocity of the jittering motion is assumed to be proportional to the frequency 
of the applied waveforms while the latter is reduced from 2.5~MHz to zero. For $^{12}$CO at low magnetic 
fields, the survival probability during deceleration is only 1/4 of the survival probability of 
$^{12}$CO guided at a constant velocity of 300~m/s. The magnetic field needed to suppress losses 
is smaller, however. While the survival probability for guiding is symmetric around a magnetic 
field of +8~Gauss, the symmetry point for deceleration is shifted closer to zero field, to 
+4~Gauss, due to the lower rotation frequency of the electric field vectors in the trap at 
lower velocities. For $^{13}$CO, the model predicts that the transmission probability during 
deceleration is larger than for guiding at all magnetic field strengths.

The differences in transmission probability between guiding and deceleration can result 
from various effects that either enhance or suppress losses as the deceleration of the 
trap increases. The smaller spatial acceptance of a strongly accelerated trap results in a 
larger fraction of the trapped cloud being in the jittering region at any given time, which 
enhances the losses during deceleration \cite{meek09}. Losses are also enhanced due to the 
molecules spending a longer time on the chip. On the other hand, since the 
average velocity of a decelerating trap is lower than that of a trap at constant velocity, 
the velocity of the jittering motion is reduced, suppressing non-adiabatic losses. While 
it is difficult to predict through simple arguments the relative importance of these effects, the 
outcome of the calculations is corroborated by previous measurements at zero magnetic field, 
in which it was shown that $^{13}$CO molecules can be decelerated to a standstill while 
$^{12}$CO molecules are rapidly lost with increasing deceleration \cite{meek09a}.

\section{Conclusions}
In this paper we have studied the losses due to non-adiabatic transitions in metastable CO
molecules --- laser-prepared in the upper $\Lambda$-doublet component of the $J=1$ level
in the $a\, ^3\Pi_1$, $v=0$ state --- guided at a constant velocity in microtraps over a chip.
Transitions between levels in which the molecules are trapped and levels in which the
molecules are not trapped can be suppressed (enhanced) when the energetic splitting
between these levels is increased (decreased) by the application of a static magnetic field.
For a quantitative understanding of this effect, the energy level structure of $^{12}$CO and
$^{13}$CO molecules in combined magnetic and electric fields has been analyzed in detail.
When the CO molecules are guided over the chip, they are in an electric field that rotates with
a constant frequency; the direction of the externally applied magnetic field is perpendicular
to the plane of the electric field. The probability with which either $^{12}$CO or $^{13}$CO
molecules are transmitted over the chip, i.e.\ the probability that the molecules stay in a trapped
level for the complete duration of the flight over the chip, has been measured as a function
of the magnetic field. The observed transmission probability can be quantitatively explained.

To reduce trap losses in future experiments, it will be important to improve the applied 
waveforms. This will not only reduce losses due to non-adiabatic transitions caused by the jittering, 
but it will also reduce losses due to mechanical heating.  Mechanical losses might also have 
been present in the current experiment, but because the measurements always compared the guiding
efficiency with magnetic field on and off, we have not been sensitive to these losses.  
Alternatively, it might be possible to avoid the need for improved waveforms by moving the minimum
on an orbit that is much larger than the amplitude of the jittering motion, creating a large region 
around the effective trap center through which the minimum never passes.  Such a trap, known as a 
time orbiting potential (TOP) trap, prevents non-adiabatic losses but is much shallower than a 
static trap \cite{petrich95}.

The intrinsic difficulties with making the waveforms required for the experiments discussed here 
should be stressed; with present day technology these waveforms can hardly be made better than we 
have them now, in particular because, in order to bring molecules to a standstill, we 
want to be able to rapidly chirp the frequency down from 2.5~MHz to zero. While the LC filter 
used to produce the ``improved waveforms'' reduces the total harmonic distortion of the amplitudes 
from 7\% to 3\%, it also makes producing a constant amplitude frequency chirp more complicated. We 
are nevertheless optimistic that the jittering can be reduced 
by another factor of two to three relative to the best waveforms that 
we have used so far. In the case of $^{12}$CO, for instance, a magnetic field of 10 Gauss, 
applied in the right direction, would then already completely avoid losses due to non-adiabatic 
transitions. With the present waveforms, trap losses can only be avoided when the applied 
magnetic fields are made sufficiently high. One should realize that there is an upper limit to 
these fields, however, as at some point transitions to the lower $\Lambda$-doublet 
components can be induced, opening up a new loss-channel.

The extreme sensitivity to the details of the applied voltages results from the fact that the 
electric field minima above the chip originate from the vectorial cancellation of rather large electric field 
terms. Design studies are in progress to find an electrode geometry that is less sensitive to 
imperfections in the applied waveforms. A modified electrode geometry is also required 
to avoid trap losses at the ends of the tubular traps. Although the ends are closed in the 
present geometry by the fringe fields of adjacent electrodes, the electric field near the 
ends has components along the long axis of the trap, presumably leading to non-adiabatic losses 
even in the presence of the offset magnetic field.

\begin{acknowledgments}
The design of the electronics by G. Heyne, V. Platschkowski and T. Vetter has been
crucial for this work. This research has been funded by the European Community's Seventh 
Framework Program FP7/2007-2013 under grant agreement 216 774, and ERC-2009-AdG 
under grant agreement 247142-MolChip. G.S. gratefully acknowledges the support of the 
Alexander von Humboldt foundation.
\end{acknowledgments}

\appendix*
\section{The $\boldsymbol{a\, ^3\Pi_1}$, $\boldsymbol{v = 0}$, $\boldsymbol{J = 1}$ Hamiltonian}\label{appham}
In this Appendix, the formalism that has been used to calculate the energies 
of the $M$-components of the $J=1$ level in the $a\, ^3\Pi_1$, $v=0$ state of 
both $^{12}$CO  and $^{13}$CO in combined, but mutually orthogonal, static 
electric and magnetic fields is presented. In the coordinate system used here, 
the magnetic field vector is oriented along the $\hat{z}$ axis and the electric 
field vector is in the $xy$ plane. In this case, the molecular Hamiltonian
is invariant under reflection in the $xy$-plane. It is thus possible to separate the 
basis states into two uncoupled sets, consisting of wavefunctions that are either
symmetric or antisymmetric under reflection in the $xy$-plane, thereby reducing 
the computational complexity. The magnetic field can only couple states of the 
same parity and the same $M_F^B$ quantum number, where $M_F^B$ is the 
projection of the total angular momentum including nuclear spin along the 
$+\hat{z}$ axis. In the case of $^{12}$CO, there is no hyperfine interaction and 
thus $F \equiv J$ and $M_F^B \equiv  M_J^B$. As the electric field vector lies 
in the plane perpendicular to the quantization axis it can only couple states of 
opposite parity with $M_F^B$ differing by $\pm 1$. The two resulting sets of 
uncoupled basis states for $^{12}$CO are given by:
\begin{enumerate}
\item M$_J^B = -1, \pm$
\item M$_J^B = 0, \mp$
\item M$_J^B = 1; \pm$
\end{enumerate}
The $+$ and $-$ sign at the end describe the parity of the basis state. All states 
with the upper (lower) sign belong to one set. For $^{13}$CO there are two
sets containing six basis states each:
\begin{enumerate}
\item $F = 3/2, M_F^B = -3/2, \pm$
\item $F = 1/2, M_F^B = -1/2, \mp$
\item $F = 3/2, M_F^B = -1/2, \mp$
\item $F = 1/2, M_F^B = 1/2, \pm$
\item $F = 3/2, M_F^B = 1/2, \pm$
\item $F = 3/2, M_F^B = 3/2, \mp$
\end{enumerate}
Again, all states with the upper (lower) parity belong to one set.

Based on this formalism and using the zero-field spectroscopic parameters and 
matrix elements given in references \cite{meek10,wicke72,gammon71,carballo88,yamamoto88,wada00,field72,warnerphd}, 
the corresponding Hamiltonian matrices can be calculated. Without loss of generality, the electric field vector is 
taken to be oriented along the $\hat{x}$ axis, i.e.\ $\vec{E} = E \hat{x}$. The Hamiltonian 
matrices for other orientations of the electric field vector in the $xy$ plane can be computed 
using the unitary transformation given in equation (\ref{HprimeH}); the matrices given here 
correspond to $\hat{H}'$ in this equation. The origin of the energy scale for each 
isotopologue is defined to be the lowest energy field-free state in the upper 
$\Lambda$-doublet component, as shown in Figure~\ref{colevel}. 

For $^{12}$CO, the matrices $\hat{H}_{\textrm{upper}}$ and $\hat{H}_{\textrm{lower}}$ 
for the set of basis states with the upper and lower parity, respectively, are given by:
\begin{equation}
\hat{H}_{\textrm{upper}} =
\begin{pmatrix}
-R_1& S& 0\\
S& -\Lambda& S\\
0& S& R_1\\
\end{pmatrix}
\end{equation}
and
\begin{equation}
\hat{H}_{\textrm{lower}} =
\begin{pmatrix}
-\Lambda - R_2& S& 0\\
S& 0& S\\
0& S& -\Lambda + R_2\\
\end{pmatrix}
\end{equation}
where
$\Lambda = 394.066~\textrm{MHz}$\\
$R_1 = \langle 1, + | \hat{H}_Z | 1, + \rangle = 0.3332\mu_B B $\\
$R_2 = \langle 1, - | \hat{H}_Z | 1, - \rangle = 0.3406\mu_B B $\\
$S = \langle 0, + | \hat{H}_S | 1, - \rangle = 0.3513\mu_E E$,\\
and the bra and ket vectors have the form $|M_J^B, \textrm{parity}\rangle$.
It is clear from these matrices, that the energy level labeled as
$M_J^B$=0 in the case of $^{12}$CO (upper right panel of Figure~\ref{colevel})
is only directly coupled to the $M_J^B$=$\pm$1 levels of the lower 
$\Lambda$-doublet component and that there is no direct coupling to the 
nearby $M_J^B$=$\pm$1 levels of the upper $\Lambda$-doublet 
component. Provided that the electric and magnetic fields are exactly
perpendicular, this $M_J^B$=0 level is therefore stable 
against non-adiabatic transitions. 

For $^{13}$CO, the corresponding matrices for the sets of basis
states with the upper and lower parity are given by:  
\begin{widetext}
\begin{equation}
\hat{H}_{\textrm{upper}} =
\begin{pmatrix}
E_1 -3 R_5& \sqrt{3} S_3& \frac{\sqrt{3}}{2} S_4& 0& 0& 0\\
\sqrt{3} S_3& -E_3 -R_2& R_4& S_1& -S_3& 0\\
\frac{\sqrt{3}}{2} S_4& R_4& -E_2 -R_6& S_2& S_4& 0\\
0& S_1& S_2& R_1& R_3& -\sqrt{3} S_2\\
0& -S_3& S_4& R_3& E_1 + R_5& \frac{\sqrt{3}}{2} S_4\\
0& 0& 0& -\sqrt{3} S_2& \frac{\sqrt{3}}{2} S_4& -E_2 + 3 R_6\\
\end{pmatrix}
\end{equation}
and
\begin{equation}
\hat{H}_{\textrm{lower}} =
\begin{pmatrix}
-E_2 -3 R_6& \sqrt{3} S_2& \frac{\sqrt{3}}{2} S_4& 0& 0& 0\\
\sqrt{3} S_2& -R_1& R_3& S_1& -S_2& 0\\
\frac{\sqrt{3}}{2} S_4& R_3& E_1 -R_5& S_3& S_4& 0\\
0& S_1& S_3& -E_3 +R_2& R_4& -\sqrt{3} S_3\\
0& -S_2& S_4& R_4& -E_2 +R_6& \frac{\sqrt{3}}{2} S_4\\
0& 0& 0& -\sqrt{3} S_3& \frac{\sqrt{3}}{2} S_4& E_1 + 3 R_5\\
\end{pmatrix}
\end{equation}
\end{widetext}
where
$E_1 = 58.412~\textrm{MHz}$\\
$E_2 = 309.340~\textrm{MHz}$\\
$E_3 = 346.346~\textrm{MHz}$\\
$R_1 = \langle 1/2, 1/2, + | \hat{H}_Z | 1/2, 1/2, + \rangle = 0.2264\mu_B B $\\
$R_2 = \langle 1/2, 1/2, - | \hat{H}_Z | 1/2, 1/2, - \rangle = 0.2309\mu_B B$\\
$R_3 = \langle 1/2, 1/2, + | \hat{H}_Z | 3/2, 1/2, + \rangle = 0.1600\mu_B B$\\
$R_4 = \langle 1/2, 1/2, - | \hat{H}_Z | 3/2, 1/2, - \rangle = 0.1635\mu_B B$\\
$R_5 = \langle 3/2, 1/2, + | \hat{H}_Z | 3/2, 1/2, + \rangle = 0.1132\mu_B B $\\
$R_6 = \langle 3/2, 1/2, - | \hat{H}_Z | 3/2, 1/2, - \rangle = 0.1155\mu_B B $\\
$S_1 = \langle 1/2, -1/2, + | \hat{H}_S | 1/2, 1/2, -\rangle = 0.3313\mu_E E$\\
$S_2 = \langle 3/2, -1/2, - | \hat{H}_S | 1/2, 1/2, + \rangle = 0.1172\mu_E E$\\
$S_3 = \langle 3/2, -1/2, + | \hat{H}_S | 1/2, 1/2, - \rangle = 0.1172\mu_E E$\\
$S_4 = \langle 3/2, -1/2, + | \hat{H}_S | 3/2, 1/2, - \rangle = 0.3313\mu_E E$,\\
and the basis vectors have the form $|F,M_F^B,\textrm{parity}\rangle$.

If the magnetic field is not perpendicular to the electric field, additional non-zero matrix 
elements will appear in the Hamiltonian that couple the states of $\hat{H}_{\textrm{upper}}$ 
and $\hat{H}_{\textrm{lower}}$.


\begin{thebibliography}{10}

\bibitem{meek09a}
S.~A. Meek, H.~Conrad, and G.~Meijer.
\newblock Trapping molecules on a chip.
\newblock {\em Science}, 324:1699--1702, 2009.

\bibitem{kirste09}
M.~Kirste, B.~G. Sartakov, M.~Schnell, and G.~Meijer.
\newblock Nonadiabatic transitions in electrostatically trapped ammonia
  molecules.
\newblock {\em Physical Review A}, 79:051401, 2009.

\bibitem{wall10}
T.~E. Wall, S.~K. Tokunaga, E.~A. Hinds, and M.~R. Tarbutt.
\newblock Nonadiabatic transitions in a {{Stark}} decelerator.
\newblock {\em Physical Review A}, 81:033414, 2010.

\bibitem{petrich95}
W.~Petrich, M.~H. Anderson, J.~R. Ensher, and E.~A. Cornell.
\newblock Stable, tightly confining magnetic trap for evaporative cooling of
  neutral atoms.
\newblock {\em Physical Review Letters}, 74:3352--3355, 1995.

\bibitem{fortagh07}
J.~Fort{\'a}gh and C.~Zimmermann.
\newblock Magnetic microtraps for ultracold atoms.
\newblock {\em Reviews of Modern Physics}, 79:235--289, 2007.

\bibitem{meek08}
S.~A. Meek, H.~L. Bethlem, H.~Conrad, and G.~Meijer.
\newblock Trapping molecules on a chip in traveling potential wells.
\newblock {\em Physical Review Letters}, 100:153003, 2008.

\bibitem{meek09}
S.~A. Meek, H.~Conrad, and G.~Meijer.
\newblock A {{Stark}} decelerator on a chip.
\newblock {\em New Journal of Physics}, 11:055024, 2009.

\bibitem{meek10}
S.~A. Meek.
\newblock {\em A {{Stark}} Decelerator on a Chip}.
\newblock PhD thesis, Freie Universit{\"a}t Berlin, 2010.

\bibitem{wicke72}
B.~G. Wicke, R.~W. Field, and W.~Klemperer.
\newblock Fine structure, dipole moment, and perturbation analysis of {{$a\,
  ^3\Pi$ CO}}.
\newblock {\em Journal of Chemical Physics}, 56:5758--5770, 1972.

\bibitem{gammon71}
R.~H. Gammon, R.~C. Stern, M.~E. Lesk, B.~G. Wicke, and W.~Klemperer.
\newblock Metastable {{$a\, ^3\Pi$ $^{13}$CO}}: Molecular-beam
  electric-resonance measurements of the fine structure, hyperfine structure,
  and dipole moment.
\newblock {\em Journal of Chemical Physics}, 54:2136--2150, 1971.

\bibitem{carballo88}
N.~Carballo, H.~E. Warner, C.~S. Gudeman, and R.~C. Woods.
\newblock The microwave spectrum of {{CO}} in the {{$a\, ^3\Pi$}} state. ii.
  the submillimeter wave transitions in the normal isotope.
\newblock {\em Journal of Chemical Physics}, 88:7273--7286, 1988.

\bibitem{yamamoto88}
S.~Yamamoto and S.~Saito.
\newblock The microwave spectra of {{CO}} in the electronically excited states
  ({{$a\, ^3\Pi_r$}} and {{$a'\, ^3\Sigma^+$}}).
\newblock {\em Journal of Chemical Physics}, 89:1936--1944, 1988.

\bibitem{wada00}
A.~Wada and H.~Kanamori.
\newblock Submillimeter-wave spectroscopy of {{CO}} in the {{$a\, ^3\Pi$}}
  state.
\newblock {\em Journal of Molecular Spectroscopy}, 200:196--202, 2000.

\bibitem{field72}
R.~W. Field, S.~G. Tilford, R.~A. Howard, and J.~D. Simmons.
\newblock Fine structure and perturbation analysis of the {{$a\, ^3\Pi$}} state
  of {{CO}}.
\newblock {\em Journal of Molecular Spectroscopy}, 44:347--382, 1972.

\bibitem{warnerphd}
H.~E. Warner.
\newblock {\em The microwave spectroscopy of ions and other transient species
  in {{DC}} glow and extended negative glow discharges}.
\newblock PhD thesis, University of Wisconsin - Madison, 1988.

\bibitem{zeldovich67}
Y.~B. Zeldovich.
\newblock The quasienergy of a quantum-mechanical system subjected to a
  periodic action.
\newblock {\em Soviet Physics JETP}, 24:1006--1008, 1967.

\bibitem{ritus67}
V.~I. Ritus.
\newblock Shift and splitting of atomic energy levels by the field of an
  electromagnetic wave.
\newblock {\em Soviet Physics JETP}, 24:1041--1044, 1967.

\end{thebibliography}

\end{document}